\documentclass{aastex631}

 \hypersetup{linkcolor=red,citecolor=blue,filecolor=cyan,urlcolor=magenta}

\usepackage{amsmath}
\usepackage{soul}

\begin{document}

\title{Unraveling the Trigger Mechanism of Explosive Reconnection in Partially Ionized Solar Plasma}

\author{Abdullah Zafar}
\affiliation{Yunnan Observatories, Chinese Academy of Science,
Kunming, Yunnan 6502016 PR china}

\author{Lei Ni}
\affiliation{Yunnan Observatories, Chinese Academy of Science,
Kunming, Yunnan 6502016 PR china} 
\affiliation{Center for Astronomical Mega-Science, Chinese Academy of Sciences, 20A Datun Road, Chaoyang District, Beijing 100012, PR China}
\affiliation{University of Chinese Academy of Sciences, Beijing 100049, PR China}
\correspondingauthor{Lei Ni}
\email{leini@ynao.ac.cn}

\author{Jun Lin}
\affiliation{Yunnan Observatories, Chinese Academy of Science,
Kunming, Yunnan 6502016 PR china}
\affiliation{Center for Astronomical Mega-Science, Chinese Academy of Sciences, 20A Datun Road, Chaoyang District, Beijing 100012, PR China}
\affiliation{University of Chinese Academy of Sciences, Beijing 100049, PR China}

\author{Ahmad Ali}
\affiliation{Pakistan Tokamak Plasma Research Institute, Islamabad 3329, Pakistan}

\begin{abstract}

Plasmoid instability is usually accounted for the onset of fast reconnection events observed in astrophysical plasmas. 
However, the measured reconnection rate from observations can be one order of magnitude higher than that derived from MHD simulations. 
In this study, we present the results of magnetic reconnection in the partially ionized low solar atmosphere based on 2.5D magnetohydrodynamics (MHD) simulations.  
The whole reconnection process covers two different fast reconnection phases. In the first phase, the slow Sweet-Parker reconnection transits to the plasmoid-mediated reconnection, and the reconnection rate reaches about 0.02. In the second phase, a faster explosive reconnection appears, with the reconnection rate reaching above 0.06. At the same time, a sharp decrease in plasma temperature and density at the principle X-point is observed which is associated with the strong radiative cooling, the ejection of hot plasma from the local reconnection region or the motion of principle X-point from hot and denser region to cool and less dense one along the narrow current sheet. 
This causes gas pressure depletion and the increasing of magnetic diffusion at the main X-point, resulting in the local Petschek-like reconnection and a violent and rapid increase in the reconnection rate. 
This study for the first time reveals a common phenomenon that the plasmoid dominated reconnection transits to an explosive faster reconnection with the rate approaching the order of 0.1 in partially ionized plasma in the MHD scale.

\end{abstract}


\section{Introduction} \label{sec:intro}
Magnetic reconnection plays a key role in variety of process in the Universe.
For example, in active galactic nuclei, it triggers highly energetic bursts in the accretion disk around the black hole~\citep{liu2002simple}.
It has a significant impact on space weather conditions~\citep{paschmann1979plasma}, and enables phenomena like sawtooth crashes and tearing instabilities in tokamaks~\citep{furth1973tearing}. 
On the Sun, magnetic reconnection process sparks many remarkable events such as Ellerman bombs (EBs)~\citep{ellerman1917solar,georgoulis2002statistics}, Ultraviolet (UV) bursts~\citep{peter2014hot,tian2016iris}, flares~\citep{yokoyama2001magnetohydrodynamic,chen2023eruption}, surges~\citep{yokoyama1995magnetic}, jets~\citep{shibata2007chromospheric} and so on.

A long-standing problem in the field of magnetic reconnection is to predict the rate at which the magnetic reconnection proceeds.
Sweet~\citep{1958IAUS....6..123S} and Parker~\citep{parker1957sweet} separately made the first attempt to describe magnetic reconnection.
The Sweet-Parker (SP) model is characterized by a long thin current sheet between oppositely directed magnetic fields having length L, and thickness $\delta \sim L/\sqrt{S}$. 
Where $S$ is the Lundquist number and is related to the Length ($L$), the Alfven speed ($V_{A}$) and the plasma resistivity ($\eta$) by an expression $ S = LV_{A}/\eta$.
The reconnection rate scales with the Lundquist number as ~ $S^{-0.5}$. 
The value of the Lundquist number is extremely large in most plasma environments in the universe (e.g., S is about $10{^6}-10^{14}$ in the solar atmosphere).
The reconnection rate predicted by the SP model is normally several orders of magnitude slower for explaining the fast energy release process in most astrophysical environments~\citep{kulsrud1998magnetic}. 
For example, the measured reconnection rate of a solar flare current sheet in the corona is about 0.01-0.1~\citep{lin2005direct,takasao2011simultaneous}, but the predicated reconnection rate by the SP model is only about 10$^{-5}$ for a typical Lundquist number of 10$^{10}$ in the solar corona.

The discrepancy between observed and SP reconnection rates has shifted attention to other models in which the reconnection rate is not strongly dependent on the Lundquist number. 
Petschek~\citep{petschek196450} proposed one such MHD model that seems capable of explaining observed reconnection events.
In contrast to the SP model, the current sheet region in the Petschek model is much shorter, and most of the incoming fluid does not pass through it but is redirected by two pairs of slow-mode shocks on each side of the current sheet. 
The reconnection rate $\gamma$ has a weaker dependence on the Lundquist number as $\gamma \sim$~1/logS. 
Therefore, fast magnetic reconnection is expected in a large Lundquist number system. 
However, numerical simulations indicate that the classical Petschek configuration cannot be sustained unless the local resistivity or viscosity is enhanced~\citep{biskamp1986magnetic,uzdensky2000two}. 


On the other hand, the classical steady-state SP and Petschek models cannot match the unstable and explosive reconnection process usually occur in the solar and astrophysical environments. 
Many observations show that plasmoid structures are formed in the reconnection current sheet in the solar atmosphere~\citep{lin2005direct,takasao2011simultaneous,li2016magnetic}.
The plasmoid is believed to correspond to magnetic island, which refers to a region of closed magnetic fields with an O-type magnetic null point located in the center in a two-dimensional plane.
Recent studies have shown that reconnection usually takes place in a fragmented current sheet due to the presence of plasmoid instabilities and can be found in a variety of plasma models, including resistive magnetohydrodynamics (MHD)~\citep{shibata2001plasmoid,huang2012distribution,ali2019effects}, Hall MHD~\citep{shepherd2010comparison,huang2011onset} and particle-in-cell (PIC) simulations~\citep{daughton2009transition,fermo2012secondary,zhu2020relativistic-I,zhu2020relativistic-II}.  
The MHD simulations demonstrated that the reconnection rate depends weakly on the Lundquist number and is always increased to a value of order~0.01~\citep{bhattacharjee2009fast,ni2015fast}. 
However, this value is still one order of magnitude smaller than the upper limit of measured reconnection rates from the observed explosive events in the solar atmosphere~\citep{de2007high,takasao2011simultaneous,moore2011solar}.
Though the observed width of the current sheet is in the MHD scale in the solar atmosphere~\citep{lin2005direct}, the reconnection rate of order 0.1 is rarely achieved in MHD simulations with an initial equilibrium state, unless the initial system is in an unstable stage~\citep{mei2012numerical}.

Plasmas in many of astrophysical environments are usually partially ionized, such as the low solar atmosphere, protoplanetary nebulae, discs around young stellar objects and the interstellar medium. 
The neutral plasma species can strongly affect the magnetic reconnection process in different ways~\citep{ni2020magnetic}. 
Previous numerical simulations prove that collisions between ions and neutrals can directly increase magnetic diffusion and the reconnection rate in the photosphere~\citep{zafar2023high,liu2023numerical}. 
The decoupling of ions and neutrals, called ambipolar diffusion, causes faster thinning of the current sheet in the case with zero guide field~\citep{brandenburg1994formation,brandenburg1995effects}, but it does not accelerate the reconnection process after plasmoid instability appears~\citep{ni2015fast}.
The ionization-recombination effect decreases the plasma pressure inside the current sheet and accelerates magnetic reconnection~\citep{leake2012multi,murtas2021coalescence}. 
But this effect is not obvious when the ionization is stronger than the recombination in the current sheet with a significant temperature increase~\citep{ni2018magnetic}. 
Moreover, recent two-fluid MHD simulations show that the ionization-recombination process stabilizes the current sheet and dampens its dynamics~\citep{murtas2022collisional}.

In this work, we numerically investigate the non-stationary magnetic reconnection process at different altitudes in a highly stratified lower solar atmosphere.
All numerical results indicate that the local unstable Petschek-like reconnection usually occurs after the plasmoid instability appears. 
More importantly, the reconnection process is sharply accelerated when the fragment current sheet with the main X-point shows a Petschek-like structure, the maximum reconnection rate always reaches above 0.06. 
These results suggest that an efficient explosive energy release process with a reconnection rate of order 0.1 can happen in the partially ionized low solar atmosphere without kinetic reconnection mechanisms.  

\section{Numerical Model}

\subsection{Single fluid MHD equations}
In this study, we performed high-resolution 2.5D MHD simulations using NIRVANA~\citep{ziegler2011semi} code. 
All species of the hydrogen-helium plasma, such as $H$, $H^{+}$, $He$, $He^{+}$, and electrons, are strongly coupled and treated as a single fluid. 
We used the following single-fluid MHD equations in our numerical experiments:
\begin{eqnarray}
\frac{\partial \rho}{\partial t} = - \nabla \cdot (\rho \mathbf{v}),
\label{eq:1}
\end{eqnarray}

\begin{eqnarray}
\begin{aligned}
\frac{\partial (\rho \mathbf{v})}{\partial t} & = - \nabla \cdot \left[\rho \mathbf{vv}+\left(p+\frac{1}{2\mu_{0}} |\mathbf{B}|^{2}\right)I-\frac{1}{\mu_{0}}\mathbf{BB}\right] \\
&+\nabla \cdot \tau_{S},
\end{aligned}
\label{eq:2}
\end{eqnarray}

\begin{eqnarray}
\frac{\partial \mathbf{B}}{\partial t} = \nabla \times (\mathbf{v} \times \mathbf{B} - \eta \nabla \times \mathbf{B} + \mathbf{E}_{AD}),
\label{eq:3}
\end{eqnarray}

\begin{eqnarray}
\begin{aligned}
    \frac{\partial e}{\partial t} &= -\nabla \cdot \left[\left(e+p+\frac{1}{2\mu_{0}} |\mathbf{B}|^{2}\right) \mathbf{v}\right] \\
    & + \nabla \cdot \left[\frac{1}{\mu_{0}} (\mathbf{v} \cdot \mathbf{B})\mathbf{B}\right] \\
    & + \nabla \cdot \left[\mathbf{v} \cdot \tau_{s} + \frac{\eta}{\mu_{0}} \mathbf{B} \times (\nabla \times \mathbf{B})\right] \\
    & - \nabla \cdot \left[\frac{1}{\mu_{0}} \mathbf{B} \times \mathbf{E}_{AD}\right] \\
    & + Q_{rad} +\mathcal{H}, 
\end{aligned}    
\label{eq:4}
\end{eqnarray}
with
\begin{eqnarray}
e = \frac{p}{\gamma -1} + \frac{1}{2} \rho |\mathbf{v}|^{2} + \frac{1}{2 \mu_{0}} |\mathbf{B}|^{2}
\label{eq:5}
\end{eqnarray}
and
\begin{eqnarray}
p = \frac{(1.1+X_{iH}+0.1X_{iHe})\rho}{1.4 m_{i}} k_{B} T,
\label{eq:6}
\end{eqnarray}
where $\rho$, $\mathbf{v}$, $p$, $\mathbf{B}$, $e$, $T$, m$_{i}$, $k_{B}$ are the plasma mass density, fluid velocity, plasma thermal pressure, magnetic field, total energy density, plasma temperature, proton mass, and Boltzmann constant, respectively.
Whereas, $X_{iH}$ represents the ionization fraction of hydrogen, while $X_{iHe}$ denotes the helium ionization fraction. 
The total helium number density is set to 10\% of that of hydrogen, and only primary helium ionization is considered. 
The adiabatic constant $\gamma$ is set to 5/3. The stress tensor is $\tau_{S} = \xi [\nabla \mathbf{v} + (\nabla \mathbf{v})^{T} - \frac{2}{3} (\nabla \cdot \mathbf{v})I]$ , where the parameter $\xi$ is the coefficient of dynamic viscosity, expressed in the units of kg m$^{-1}$ s$^{-1}$. 
The current sheet considered in this study is parallel to the surface of the Sun, and the current sheet width thins down below 10 km during the main reconnection process, hence the gravity effect is ignored, and the initial plasma density and temperature are assumed to be uniform in the whole simulation domain.


\subsection{Diffusions and viscosity coefficients}
In the partially ionized plasma of the lower solar atmosphere, interactions between various plasma species are of key interest~\citep{wargnier2023multifluid}.
Magnetic diffusion ($\eta$) is defined as the sum of diffusion caused by electron-ion collision ($\eta_{ei}$) and electron-neutral collision ($\eta_{en}$), and is given as~\citep{khomenko2012heating,ni2022plausibility}
\begin{eqnarray}
\eta = \eta_{ei} + \eta_{en} = \frac{m_{e} \nu_{ei}}{e^{2}_{c} n_{e} \mu_{0}} + \frac{m_{e} \nu_{en}}{e^{2}_{c} n_{e} \mu_{0}}
\label{eq:7}
\end{eqnarray}
where $m_{e}$ is the mass of electron, $e_{c}$ is the charge on electron, $\mu_{0}$ is the permeability of free space, $n_{e}$ is the electron number density, and $\nu_{en}$, $\nu_{ei}$ describe the collision frequencies of electron-neutral and electron-ion, respectively. The number density of electron and collision frequencies are~\citep{ni2022plausibility}
\begin{eqnarray}
n_{e} = \frac{\rho(X_{iH}+0.1X_{iHe})}{1.4m_{i}}  
\label{eq:8}
\end{eqnarray}

\begin{eqnarray}
\nu_{ei} = \frac{n_{e} e^{4}_{c} \Lambda}{3 m^{2}_{e} \epsilon^{2}_{0}} \left( \frac{m_{e}}{2 \pi k_{B} T} \right)^{3/2},
\label{eq:9}
\end{eqnarray}
and 
\begin{eqnarray}
\nu_{en} = n_{n} \sqrt{\frac{8 k_{B} T}{\pi m_{en}}} \sigma_{en}.
\label{eq:10}
\end{eqnarray}
where $\Lambda$, $\epsilon_{0}$, $n_{n}$, $\sigma_{en}$ denote the Coulomb logarithm, vacuum permittivity, neutrals number density, and collisional cross-section, respectively. The Coulomb logarithm is expressed as~\citep{ni2022plausibility}
\begin{eqnarray}
\Lambda = 23.4-1.15 \log_{10} n_{e} + 3.45 \log_{10} T.
\label{eq:11}
\end{eqnarray}
In the hydrogen-helium mixture, the collision frequency between electrons and neutrals ($\nu_{en}$) is contributed by both the collisions of electrons with neutral helium and neutral hydrogen. The $\nu_{en}$ is

\begin{eqnarray}
\nu_{en} = n_{n H_{e}} \sqrt{\frac{8 k_{B} T}{\pi m_{e}}} \sigma_{e-n H_{e}} + n_{n H} \sqrt{\frac{8 k_{B} T}{\pi m_{e}}} \sigma_{e-n H},
\label{eq:12}
\end{eqnarray}
where $n_{nHe} = 0.1\rho(1-Y_{iHe})/(1.4mi)$, $n_{nH} = \rho(1-Y_{iH})/(1.4mi)$ are the number densities of neutral helium and the neutral hydrogen, respectively. In our simulations, the electron-neutral hydrogen collision cross-section is $2 \times 10^{-19}$ m$^{2}$, whereas electron-neutral helium collision cross-sections is $\sigma_{e-n H}/3$~\citep{vranjes2013collisions}. After simplification, the magnetic diffusivities becomes
\begin{eqnarray}
\eta_{ei} \simeq 1.0246 \times 10^{8} \Lambda T^{-1.5}
\label{eq:13}
\end{eqnarray}
and
\begin{eqnarray}
\eta_{en} \simeq 0.0351 \sqrt{T} \frac{\left[\frac{0.1}{3} (1-X_{iH_{e}})+(1-X_{iH})\right]}{X_{iH} + 0.1 X_{iHe}}
\label{eq:14}
\end{eqnarray}
which are in units of m$^{2}$ s$^{-1}$.


The ambipolar electric field (E$_{AD}$) in the induction equation~(Eq.~(\ref{eq:3})) and the energy equation~(Eq.~(\ref{eq:4})) is~\citep{ni2021magnetic,ni2022plausibility}
\begin{eqnarray}
\mathbf{E}_{AD} = \frac{1}{\mu_{0}} \eta_{AD} [(\nabla \times B) \times B ] \times B,
\label{eq:15}
\end{eqnarray}
where $\eta_{AD}$ represents the coefficient of the ambipolar diffusion and is expressed as~\citep{khomenko2012heating,ni2020magnetic,ni2022plausibility} 
\begin{eqnarray}
\eta_{AD} = \frac{(\rho_{n}/\rho)^2}{\rho_{i} \nu_{in} + \rho_{e} \nu_{en}}
\label{eq:16}
\end{eqnarray}
the ambipolar diffusion coefficient is measured in m$^{3}$ s kg$^{-1}$ units. For hydrogen-helium plasma, the ratio of the mass density of neutral species to total mass density is~\citep{ni2022plausibility}
\begin{eqnarray}
\rho_{n}/\rho  = \frac{0.4 (1- X_{iHe}) + (1- X_{iH})}{1.4}.
\label{eq:17}
\end{eqnarray}
The ion collision part is written as~\citep{ni2022plausibility}
\begin{eqnarray}
\begin{aligned}
    \rho_{i} \nu_{in} & = \rho_{iH} n_{nH} \sqrt{\frac{8 k_{B} T}{\pi m_{i}/2}} \sigma_{iH-nH} \\
    & + \rho_{iH} n_{nHe} \sqrt{\frac{8 k_{B} T}{4 \pi m_{i}/5}} \sigma_{iH-nHe} \\
    & + \rho_{iHe} n_{nH} \sqrt{\frac{8 k_{B} T}{4 \pi m_{i}/5}} \sigma_{iHe-nH} \\
    & + \rho_{iHe} n_{nHe} \sqrt{\frac{8 k_{B} T}{2 \pi m_{i}}} \sigma_{iHe-nHe}, 
\end{aligned}
\label{eq:18}
\end{eqnarray}
where $\rho_{iH} = \rho X_{iH}/1.4$, $\rho_{iHe} = 0.4\rho X_{iHe}/1.4$ are the mass densities of ionized hydrogen and helium, respectively; $\sigma_{iH-nH}$ represents the collision cross-section of ionized hydrogen and neutral hydrogen, $\sigma_{iH-nHe}$ is the collisional cross-section between ionized hydrogen and neutral helium, $\sigma_{iHe-nH}$ is the cross-section for the ionized helium and neutral hydrogen collision, and  $\sigma_{iHe-nHe}$ corresponds to that of the ionized helium-neutral helium collision cross-section. We consider $\sigma_{iH-nH} = 1.5 \times 10^{-18}$ m$^{2}$, $\sigma_{iH-nHe}$ = $\sigma_{iHe-nH}$ = $\sigma_{iHe-nHe}$ = $\sigma_{iH-nH}/\sqrt{3}$~\citep{vranjes2013collisions,barata2010elastic}. 
The electron collision part reads~\citep{ni2022plausibility} 
\begin{eqnarray}
\begin{aligned}
    \rho_{e} \nu_{en} & = \rho_{e} n_{nH} \sqrt{\frac{8 k_{B} T}{\pi m_{e}}} \sigma_{e-nH} \\
    & + \rho_{e} n_{nHe} \sqrt{\frac{8 k_{B} T}{\pi m_{e}}} \sigma_{e-nHe}, \\
\end{aligned}
\label{eq:19}
\end{eqnarray}
where $\sigma_{e-nH}$, $\sigma_{e-nHe}$ are the collisional cross-sections of electrons with neutral hydrogen and neutral helium species, respectively. The collisional cross-sections contributed by electrons and neutrals are smaller than the collisional cross-section between ions and neutrals and are ignored. The coefficient of ambipolar diffusion (Eq.~(\ref{eq:16})) is simplified into $\eta_{AD} = \frac{(\rho_{n}/\rho)^2}{\rho_{i} \nu_{in}}$.


In partially ionized plasma, the dynamic viscosity is contributed by both ions and neutrals, and is given by~\citep{ni2022plausibility}
\begin{eqnarray}
\xi = \xi_{i} + \xi_{n} = \frac{n_{i} k_{B} T}{\nu_{ii}} + \frac{n_{n} k_{B} T}{\nu_{nn}},
\label{eq:20}
\end{eqnarray}
where $\xi_{i}$, $\xi_{n}$ are the viscosity coefficients contributed by ion and neutrals, respectively, while $\nu_{nn,ii}$ corresponds to the collision frequencies between similar species, such as collisions between neutral-neutral and ion-ion. The collision frequencies between the similar species are as below~\citep{leake2013magnetic}: 
\begin{eqnarray}
\nu_{nn} = n_{n} \sigma_{nn} \sqrt{\frac{16 k_{B} T}{\pi m_{n}}}
\label{eq:21}
\end{eqnarray}
and
\begin{eqnarray}
\nu_{ii} = \frac{n_{i} e^{4}_{c} \Lambda}{3 m^{2}_{i} \epsilon^{2}_{0}} \left(\frac{m_{i}}{2 \pi k_{B} T} \right)^{3/2}.
\label{eq:22}
\end{eqnarray}
The contributions of hydrogen and helium are both considered when calculating the dynamic viscosity, which is assumed to be more realistic than the one used in previous study~\citep{ni2022plausibility}. The final form of dynamic viscosity used here is
\begin{eqnarray}
\xi = \xi_{i} + \xi_{n} \simeq  \frac{4.8692 \times 10^{-16}}{\Lambda} T^{2} \sqrt{T}+ 2.0127 \times 10^{-7} \sqrt{T}. 
\label{eq:24}
\end{eqnarray}


In our simulations, we use time-dependent ionization degrees for hydrogen and helium. The ionization degree of helium varies exponentially with temperature, X$_{iHe} = 1-10^{0.325571-0.0000596T}$. This expression is valid for both the photosphere and chromosphere layers, but it gives a negative value when the temperature is below 5413K. Therefore, we consider X$_{iHe} = 0.00010084814$ when the temperature is less than that critical value.
On the other hand, different approaches are employed to calculate the ionization degree of hydrogen (X$_{iH}$) in the photosphere and chromosphere regions, respectively.   
The modified Saha and Boltzamann equation~\citep{gan1990hydrodynamic,fang2002magnetic} is employed to calculate hydrogen ionization degree in the photosphere, which depends on plasma density and temperature. 
The temperature dependent hydrogen ionization degree for the chromospheric environment is adopted from Carlsson \& Leenaarts work~\citep{carlsson2012approximations}. 
We refer interested readers to see Figure 1 of Ni et al. 2022~\citep{ni2022plausibility} and Liu et al. 2023~\citep{liu2023numerical}, which show temperature-dependent faction of hydrogen and helium.
At the beginning, the uniform initial plasma density and temperature profiles produce a constant value of ionization degrees across the simulation domain; however, later on, these ionization degrees are updated based on the local plasma parameters.


\subsection{Radiative Cooling models}
Radiative transfer processes, or the energy-exchange processes between the solar atmosphere and the radiation field, have a significant impact on the properties and behavior of the plasma especially in the low solar atmosphere.
In order to make our numerical experiments more realistic and better understand magnetic reconnection in the lower solar atmosphere, we included different radiative cooling models in the photosphere and chromosphere layers.


For effective radiative cooling process in the photosphere layer, the Gan \& Fang~\citep{gan1990hydrodynamic} is applied in the photospheric magnetic reconnection process, read as
\begin{eqnarray}
Q_{rad1} = - 1.547 \times 10^{-42} n_{e} n_{H} \alpha T^{1.5},
\label{eq:25}
\end{eqnarray}
where n$_{e}$, n$_{H}$ are the electron and hydrogen number densities, respectively. The modified Saha and Boltzamann formula ~\citep{gan1990hydrodynamic,fang2002magnetic} is used to calculate the number density of electron. The height dependent function $\alpha$ is given by ~\citep{gan1990hydrodynamic}

\begin{eqnarray}
\begin{aligned}
\alpha &= 10^{a1} + 2.3738\times 10^{-4}e^{a2},\\
a1 &= 2.75 \times 10^{-3}Z-5.445, \\
a2 &= \frac{-Z}{163}
\label{eq:25a}
\end{aligned}
\end{eqnarray}
where $Z$ represents height in km.

The second radiative cooling model used in the chromospheric simulations of this study is the model proposed by Carlsson \& Leenaarts~\citep{carlsson2012approximations}, which is regarded as the widely acceptable model for the chromosphere. This model focuses on the effects of radiative cooling in the three chromospheric spectral lines, such as neutral hydrogen, singly ionized calcium, and singly ionized magnesium. The model is expressed as below

\begin{eqnarray}
Q_{rad2} = - \sum_{X = H, Mg, Ca} L_{Xm} (T) E_{Xm} (\tau) \frac{N_{Xm}}{N_{X}} (T) A_{X} \frac{N_{H}}{\rho} n_{e} \rho,
\label{eq:27}
\end{eqnarray}
where $L_{Xm} (T)$ is the optically thin radiative loss function that varies with temperature, per electron, and per particle of element $X$ in the ionization stage $m$, 
$E_{Xm} (\tau)$ denotes the escape probability that depends on optical depth $\tau$, 
$\frac{N_{Xm}}{N_{X}}(T)$ represents the fraction of element $X$ in the ionization stage $m$, 
$A_{X}$ denotes the element abundance $X$, 
and $\frac{N_{H}}{\rho} = 4.407 \times 10^{23} g^{-1}$ shows the number of hydrogen particles per unit mass of the chromospheric material. 
The $\tau$ in the above expression (eq.~\ref{eq:27}) is calculated by multiplying the total column density of neutral hydrogen with a constant number of $4.0 \times 10^{-14}$ cm$^{2}$.
The RADYN~\citep{1997ApJ...481..500C,2002ApJ...572..626C} code computed $L_{Xm}$, $E_{Xm}$, and $N_{Xm}/N_{X}$ for hydrogen in a 1D radiation hydrodynamic simulation with non-equilibrium ionization.
However, the components contributed by Mg and Ca were obtained from a 2D radiation-MHD simulation with BIFROST, that gave the atmospheric structure and radiative transfer calculations employing MULTI3D~\citep{leenaarts2009multi3d}.


\subsection{Simulation setup} \label{sec:setup}
In this study, we carried out high-resolution 2.5D MHD simulations to investigate the magnetic reconnection process at different altitude in the low solar atmosphere with an anti-parallel magnetic field configuration.
The initial magnetic fields of the Harris current sheet are given by $B_{x0} = -b_{0} tanh[y/(0.05L_{0})]$, $B_{y0} = 0$ and $B_{z0} = b_{0}/cosh[y/(0.05L_{0})]$ ($b_{0}$, $L_{0} = 2\times10^{5} m$ are the characteristics of magnetic field strength and system scale length, respectively).
The size of the simulation box is from $0$ to $L_{0}$ along the x-direction and $-0.5L_{0}$ to $0.5L_{0}$ along the y-direction which is large enough to allow nonlinear evolution and motion of plasmoids.
The small initial magnetic perturbations that trigger the magnetic reconnection are $b_{x1} = -b_{pert} sin[2\pi(y+0.5L_{0})/L_{0}]cos[2\pi(x+0.5L_{0})/L_{0}]$ and  $b_{y1} = b_{pert} cos[2\pi(y+0.5L_{0})/L_{0}]sin[2\pi(x+0.5L_{0})/L_{0}]$. 
Open boundary conditions are used at both directions in all simulation cases.
The adaptive mesh refinement (AMR) technique with the highest level of 9 and a base-level grid of 192 $\times$ 192 are applied in this study. The initial plasma density ($\rho_{0}$) and temperature (T$_{0}$) given in Table~\ref{tab:table1} are taken from C7 model~\citep{avrett2008models}.
The parameter Z in Table~\ref{tab:table1} represent the height above the solar surface in kilometers. The simulation at 400 km corresponds to the photosphere layer of the Sun. The cases at Z = 600 and 800 km relate to the lower chromosphere. The heights from 1000 km to 1400 km correspond to the middle chromosphere, while the 1700 km to 2000 km range represents the upper chromosphere region. 
The simulations in the middle chromosphere (Z = 1250 and 1400 km) are performed with different initial magnetic perturbations (b$_{pert}$) and plasma $\beta$ ($\beta_{0})$. 
In our simulations, we changed $\beta_{0}$ by varying the strength of the initial magnetic field.

\begin{table}[b]
\caption{\label{tab:table1}%
Input parameters used in the simulations. }\begin{ruledtabular}
\begin{tabular}{ccc}
Z (km)&
\multicolumn{1}{c}{\textrm{$\rho_{0}$(kgm$^{-3}$)}}&
\multicolumn{1}{c}{\textrm{T$_{0}$ (K)}}\\
\hline
400& \mbox{1.56$\times 10^{-5}$} & 4590 \\
600& \mbox{2.40$\times 10^{-6}$} & 4421 \\
800& \mbox{3.44$\times 10^{-7}$} & 5100 \\
1000& \mbox{6.32$\times 10^{-8}$} & 6223 \\
1250& \mbox{1.16$\times 10^{-8}$} & 6588 \\
1400& \mbox{4.51$\times 10^{-09}$} & 6610 \\
1700& \mbox{7.44$\times 10^{-10}$} & 6641 \\
2000& \mbox{1.68$\times 10^{-10}$} & 6678 \\
\end{tabular}
\end{ruledtabular}
\end{table}

\begin{figure*}[hbt!]
\centering
\begin{minipage}{0.6\textwidth}
\includegraphics[width=1.0\textwidth]{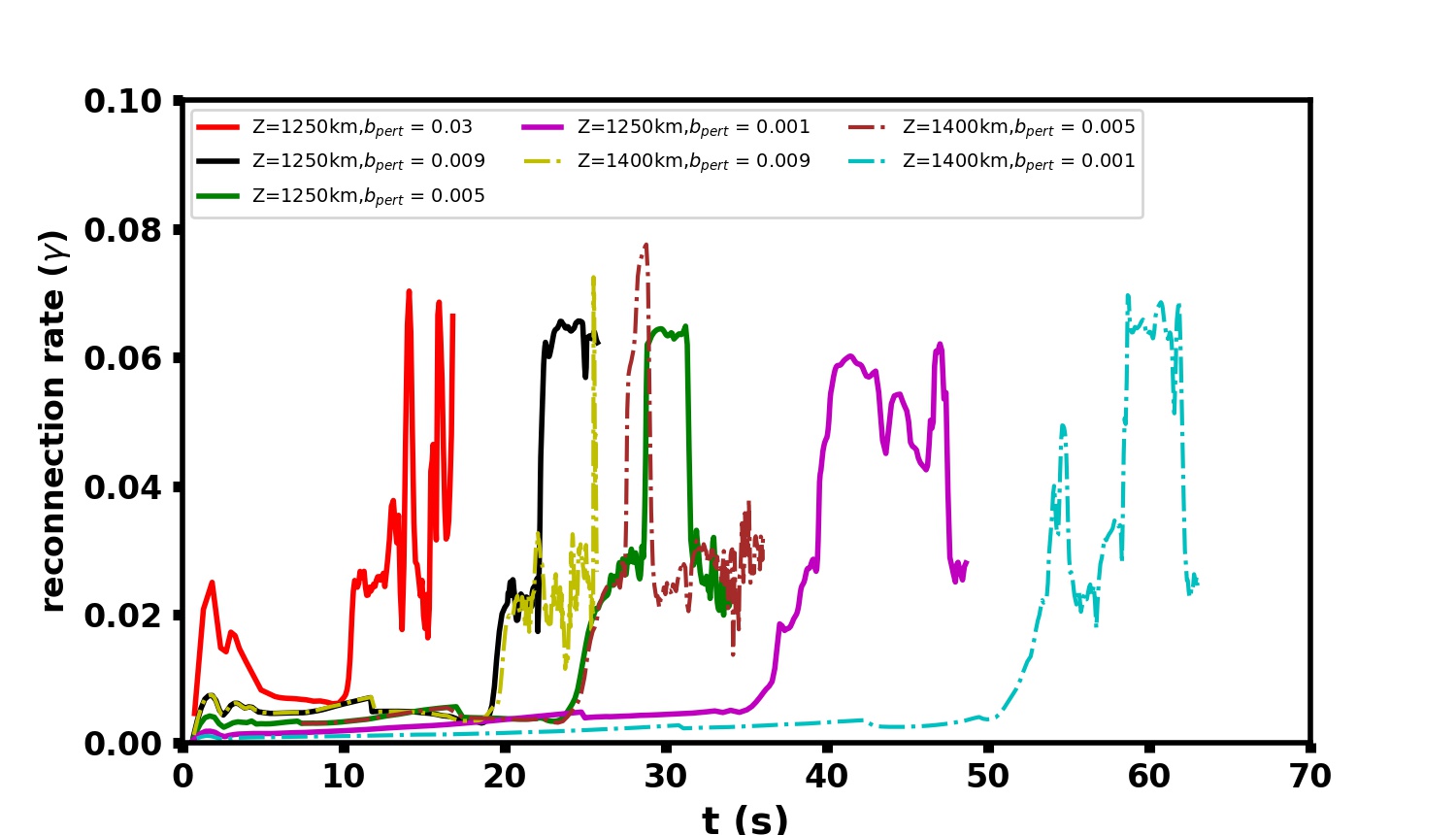}
\put(-265,120){\textbf{(a)}}
\end{minipage}
\begin{minipage}{0.6\textwidth}
\includegraphics[width=1.0\textwidth]{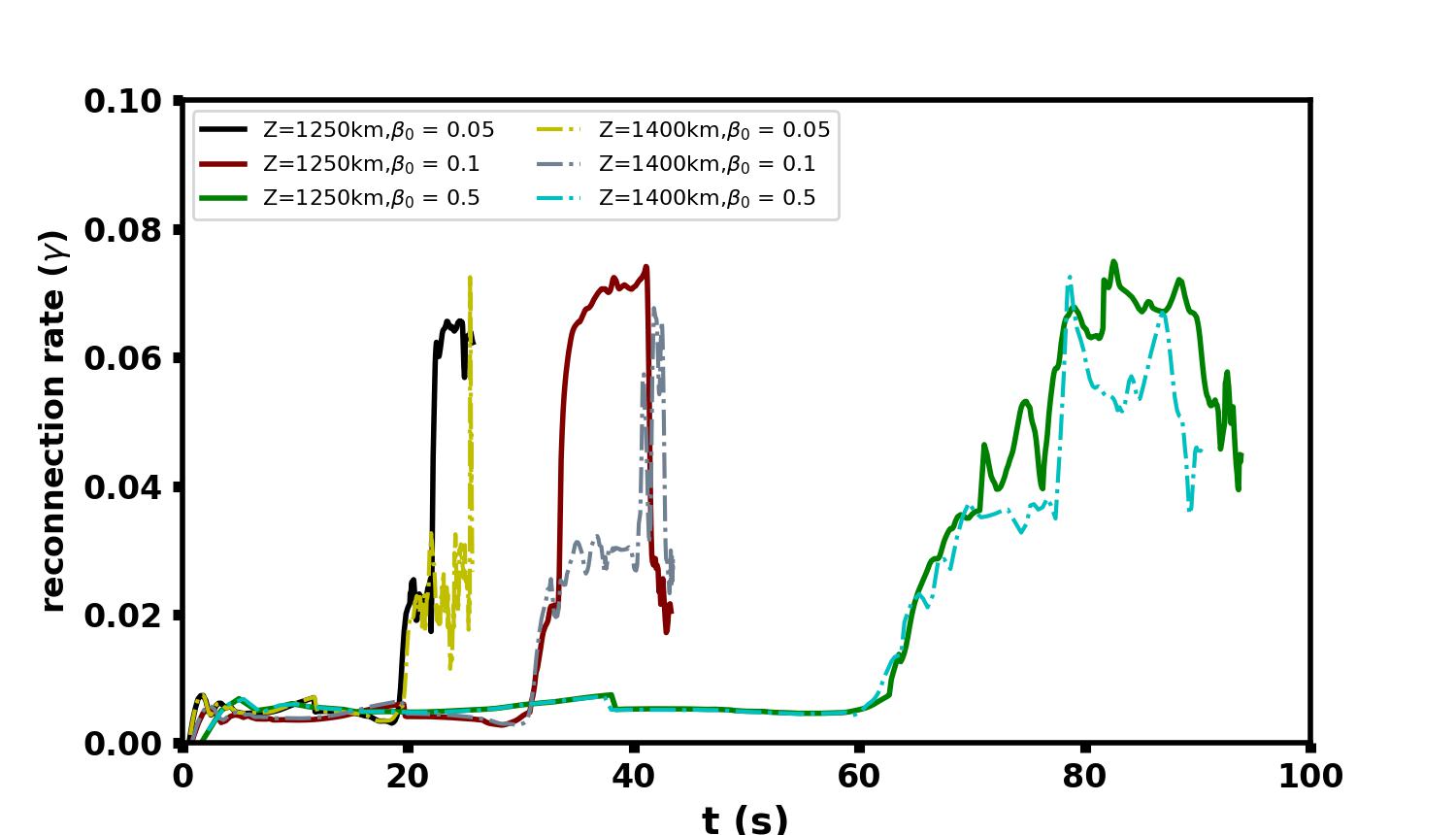}
\put(-265,120){\textbf{(b)}}
\end{minipage}
\begin{minipage}{0.62\textwidth}
\includegraphics[width=1.0\textwidth]{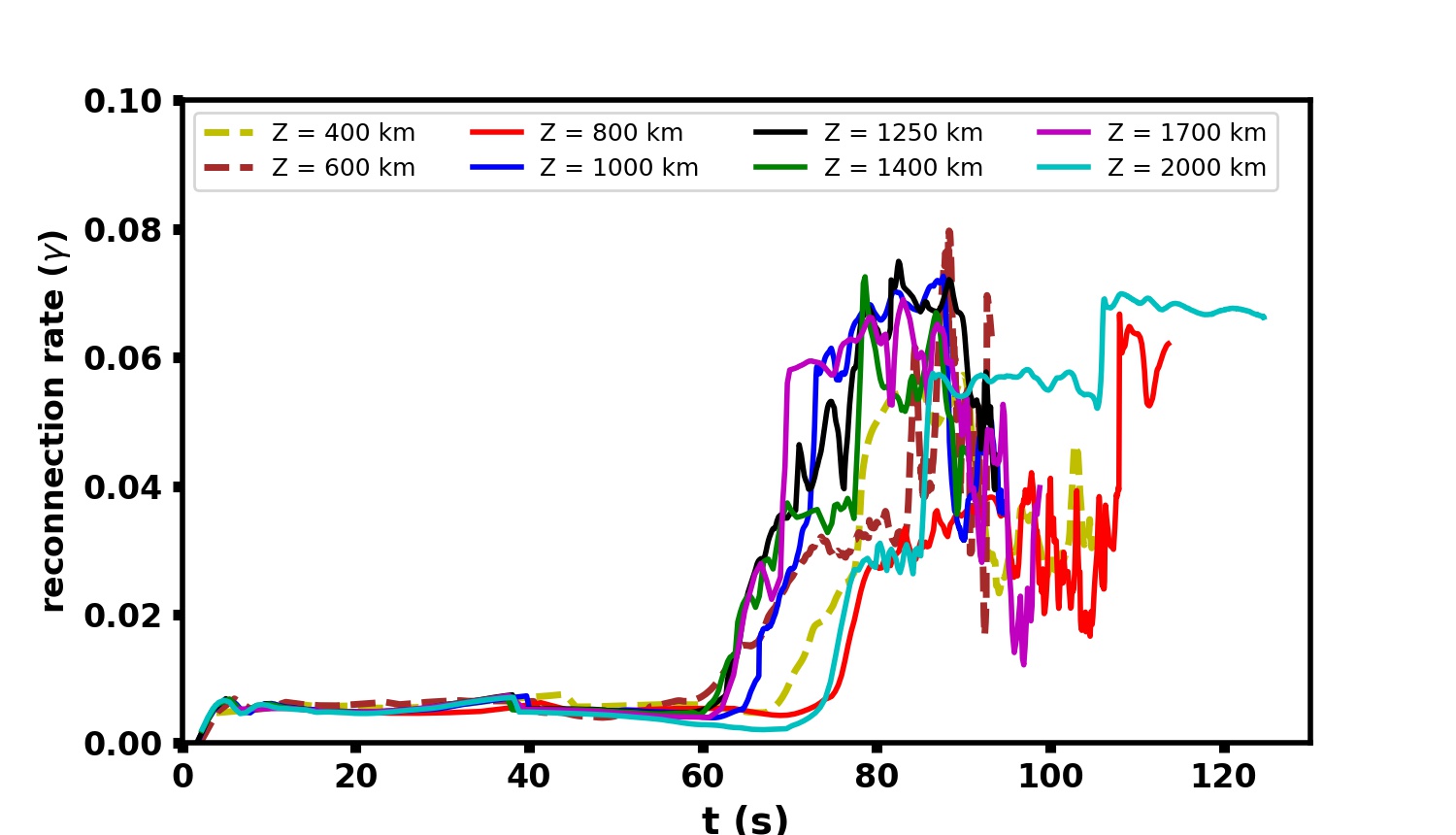}
\put(-275,130){\textbf{(c)}}
\end{minipage}
\caption{Time evolution of the reconnection rate $\gamma$ for different initial perturbations $b_{pert}$ at Z = 1250 km (solid lines) and Z = 1400 km (dash-dotted lines) with $\beta_{0}$ = 0.05 (a), for different vales of initial plasma $\beta$ at Z = 1250 km (solid lines) and Z = 1400 km (dash-dotted lines) with b$_{pert}$ = 0.009 (b) and at different altitudes in the low solar atmosphere using the same initial plasma $\beta$ ($\beta_{0}$ = 0.5) and initial perturbation ($b_{pert}$ = 0.009). The dash lines correspond to photospheric reconnection cases (c).}
\label{fig_1}
\end{figure*}

\begin{figure*}[hbt!]
\centering
\begin{minipage}{0.70\textwidth}
\includegraphics[width=1.0\textwidth]{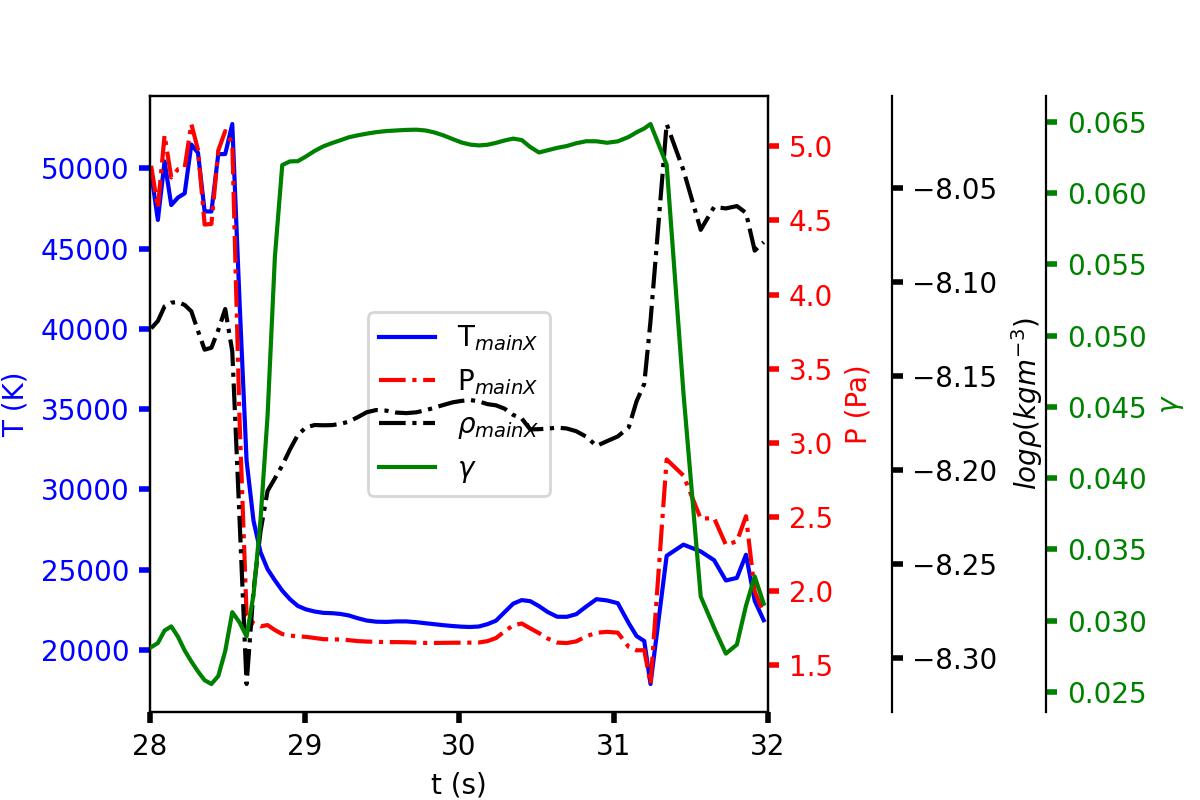}
\put(-230,150){\textbf{(a)}}
\end{minipage}

\begin{minipage}{0.45\textwidth}
\includegraphics[width=1.0\textwidth]{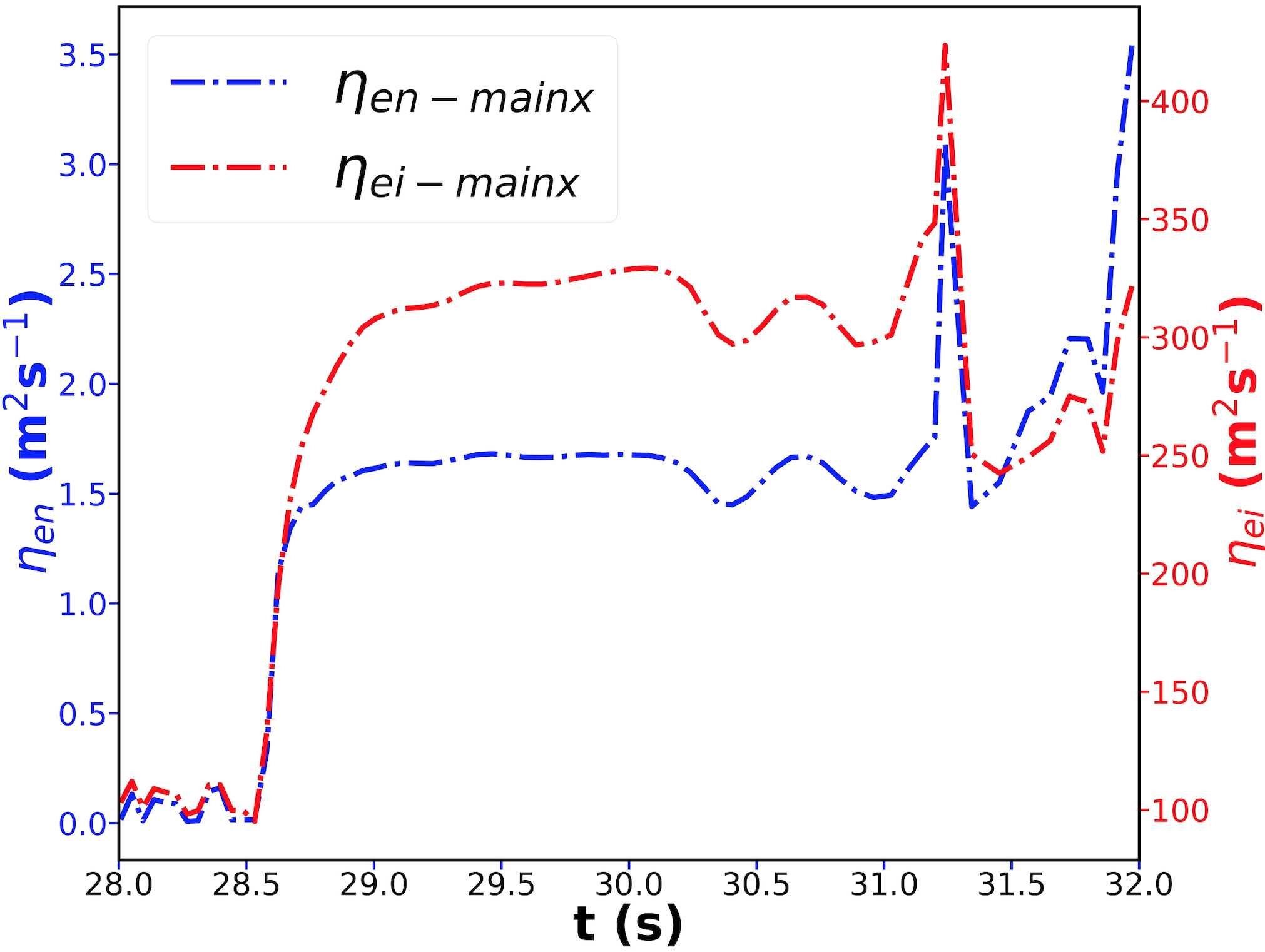}
\put(-45,30){\textbf{(b)}}
\end{minipage}
\begin{minipage}{0.45\textwidth}
\includegraphics[width=1.0\textwidth]{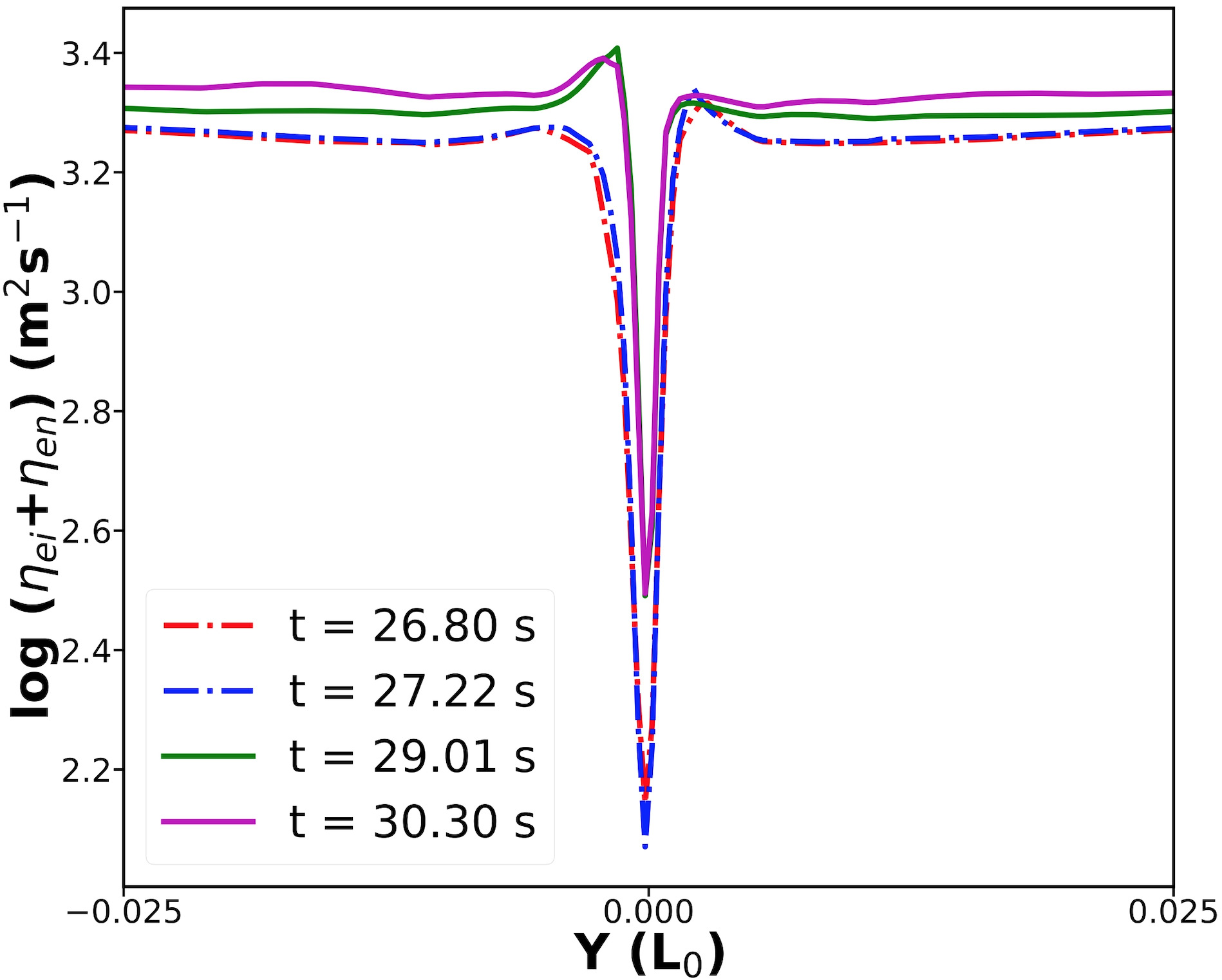}
\put(-45,30){\textbf{(c)}}
\end{minipage}
\caption{Temporal evolutions of the temperature, pressure, density at the main X-point and the reconnection rate during the explosive fast reconnection phase (a), diffusion coefficients $\eta_{en}$ and $\eta_{ei}$ at the main X-point in the newly observed explosive reconnection stage (b) and the profiles of the logarithm of the total magnetic diffusion ($\eta_{en}$+$\eta_{ei}$) at various times along the y-direction through the main X-point  in the case with $\beta_{0}$ = 0.05 and $b_{pert}$ = 0.005 at Z=1250 km above the solar surface.}
\label{fig_2}
\end{figure*}

\begin{figure*}[hbt!]
\centering
\includegraphics[width=0.60\textwidth]{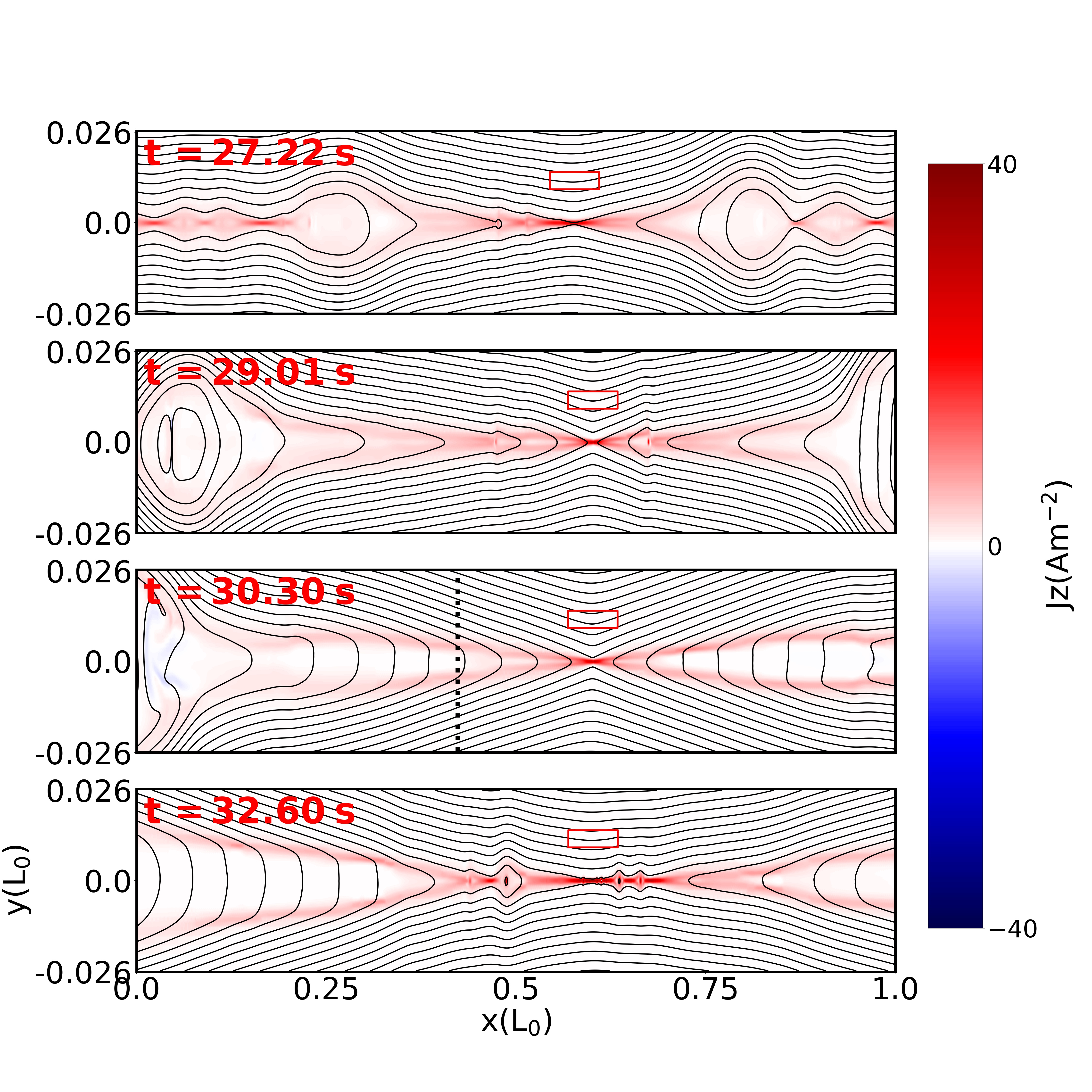}
\put(-160,275){\textbf{(a)}}
\put(-70,260){\textbf{(i)}}
\put(-70,200){\textbf{(ii)}}
\put(-75,135){\textbf{(iii)}}
\put(-75,75){\textbf{(iv)}}

\begin{minipage}{0.30\textwidth}
\includegraphics[width=1.0\textwidth]{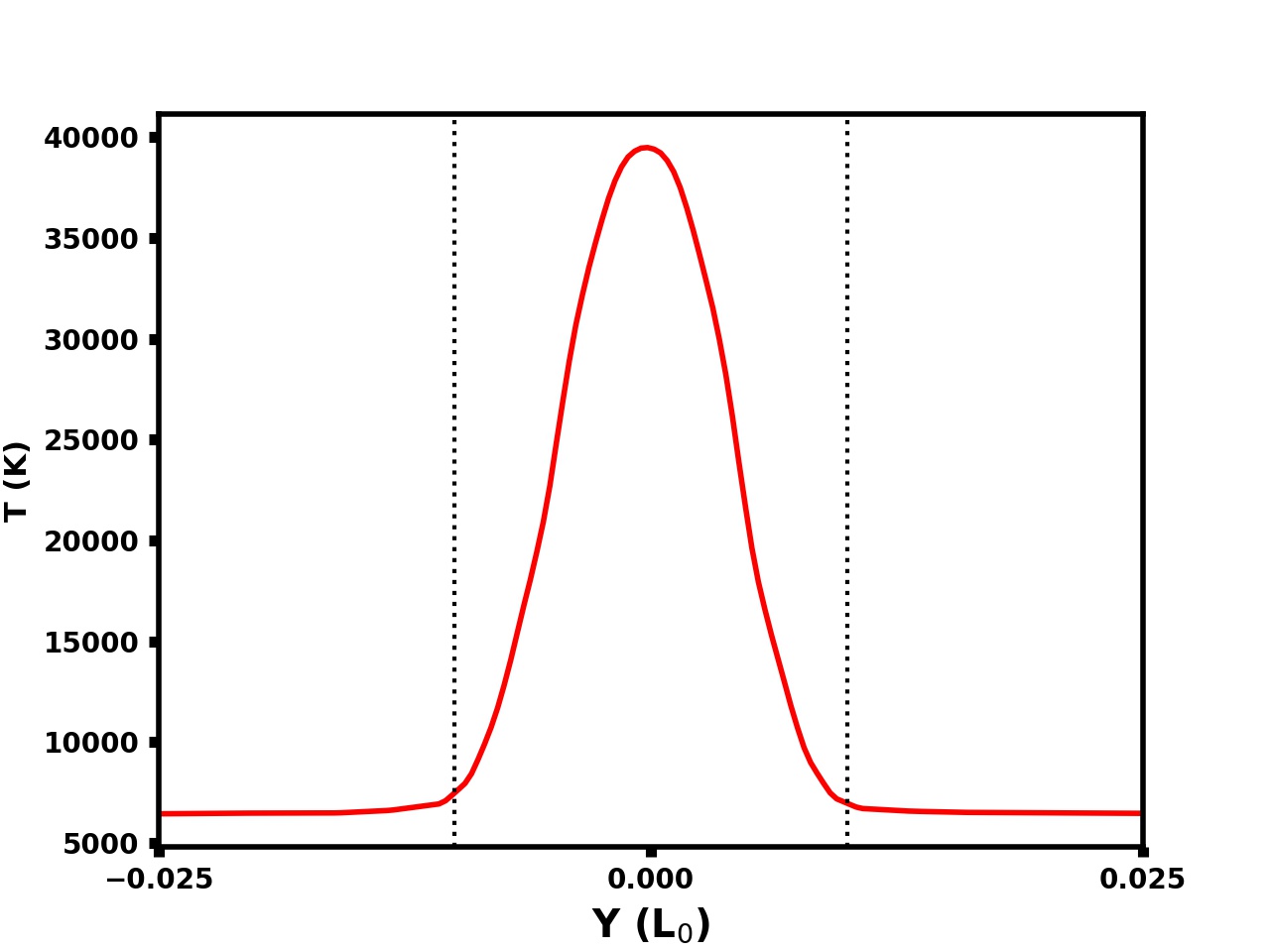}
\put(-33,23){\textbf{(b)}}
\end{minipage}
\begin{minipage}{0.30\textwidth}
\includegraphics[width=1.0\textwidth]{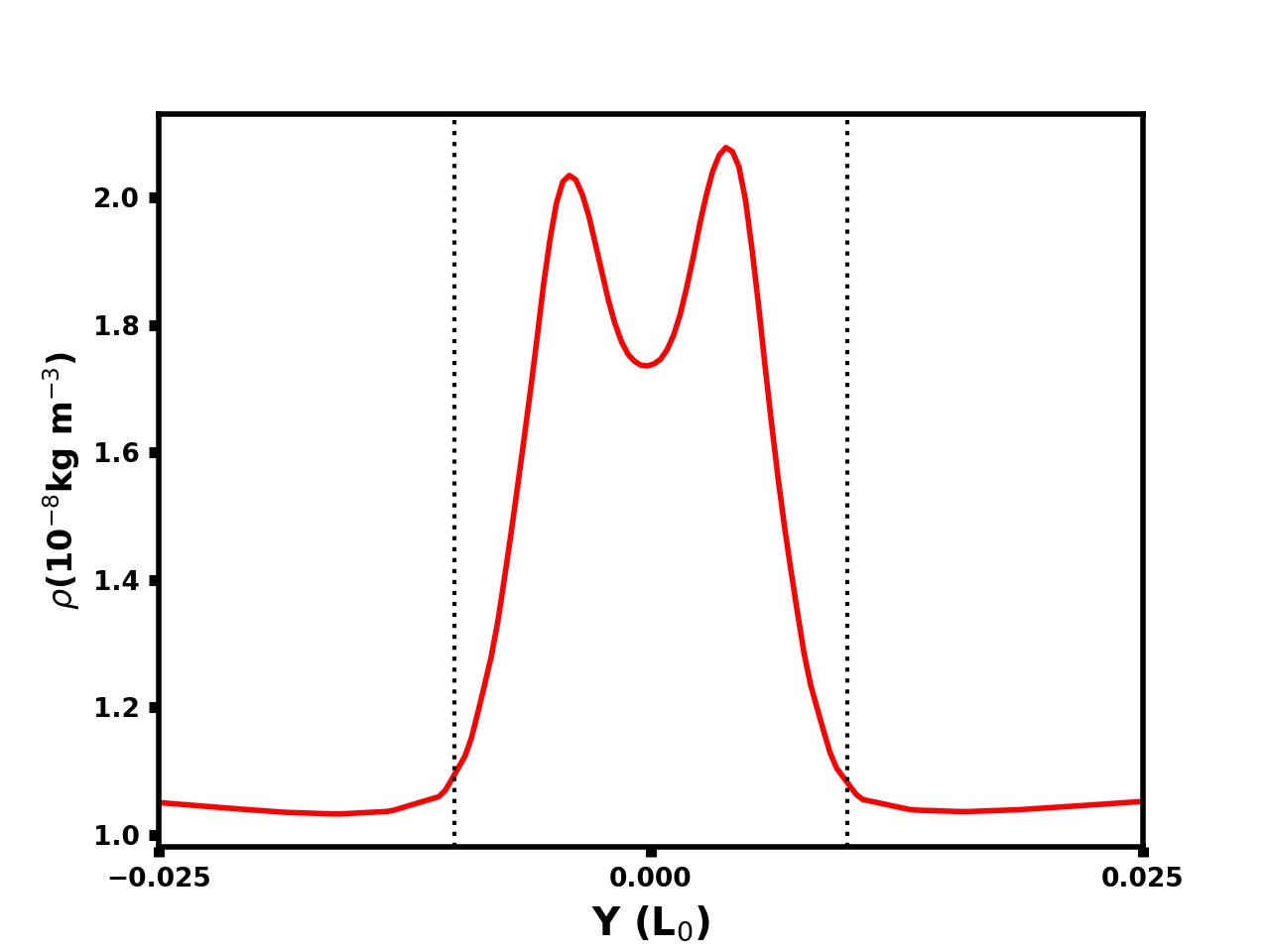}
\put(-33,23){\textbf{(c)}}
\end{minipage}
\begin{minipage}{0.30\textwidth}
\includegraphics[width=1.0\textwidth]{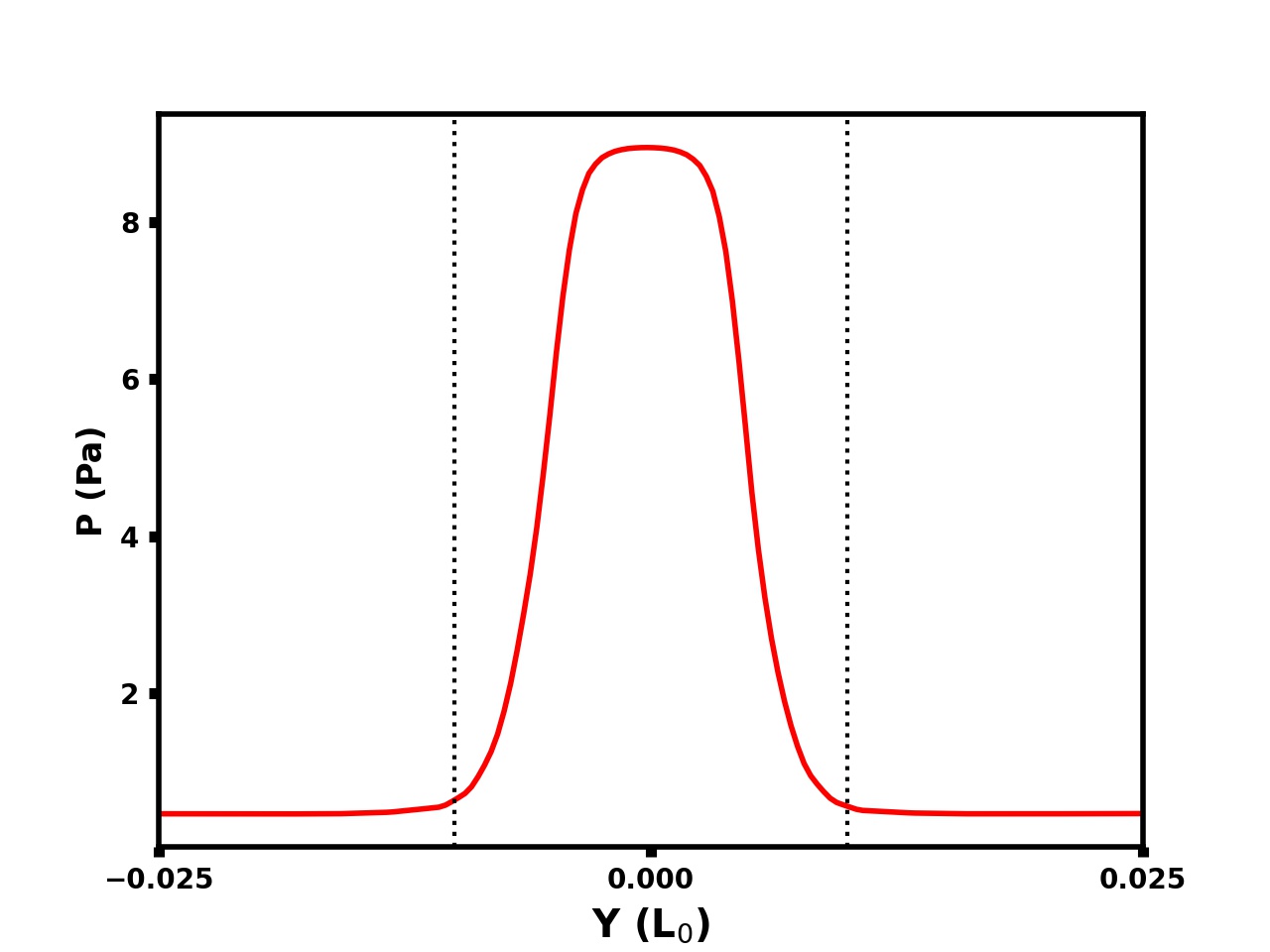}
\put(-33,23){\textbf{(d)}}
\end{minipage}

\begin{minipage}{0.30\textwidth}
\includegraphics[width=1.0\textwidth]{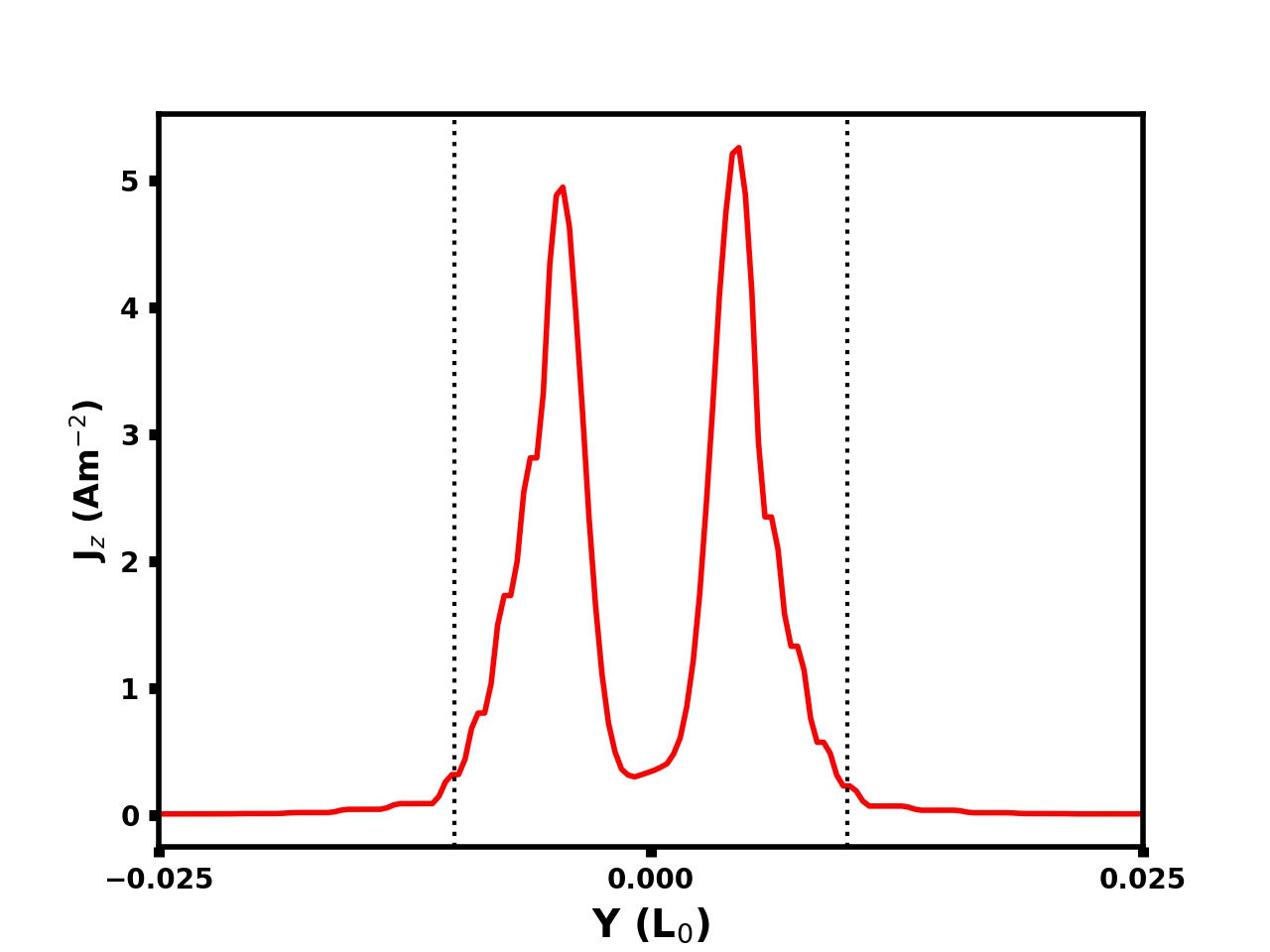}
\put(-33,23){\textbf{(e)}}
\end{minipage}
\begin{minipage}{0.30\textwidth}
\includegraphics[width=1.0\textwidth]{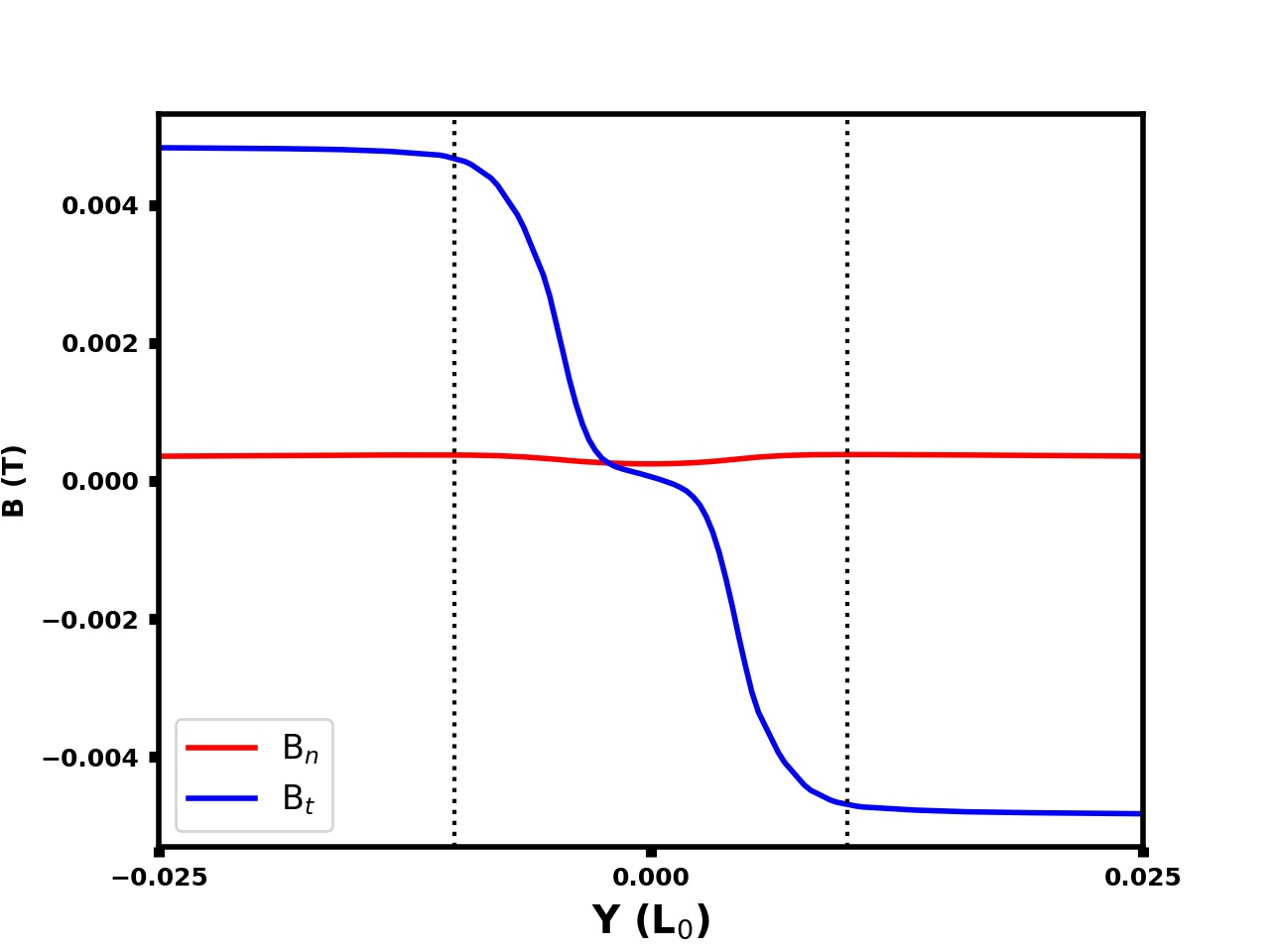}
\put(-33,23){\textbf{(f)}}
\end{minipage}
\begin{minipage}{0.30\textwidth}
\includegraphics[width=1.0\textwidth]{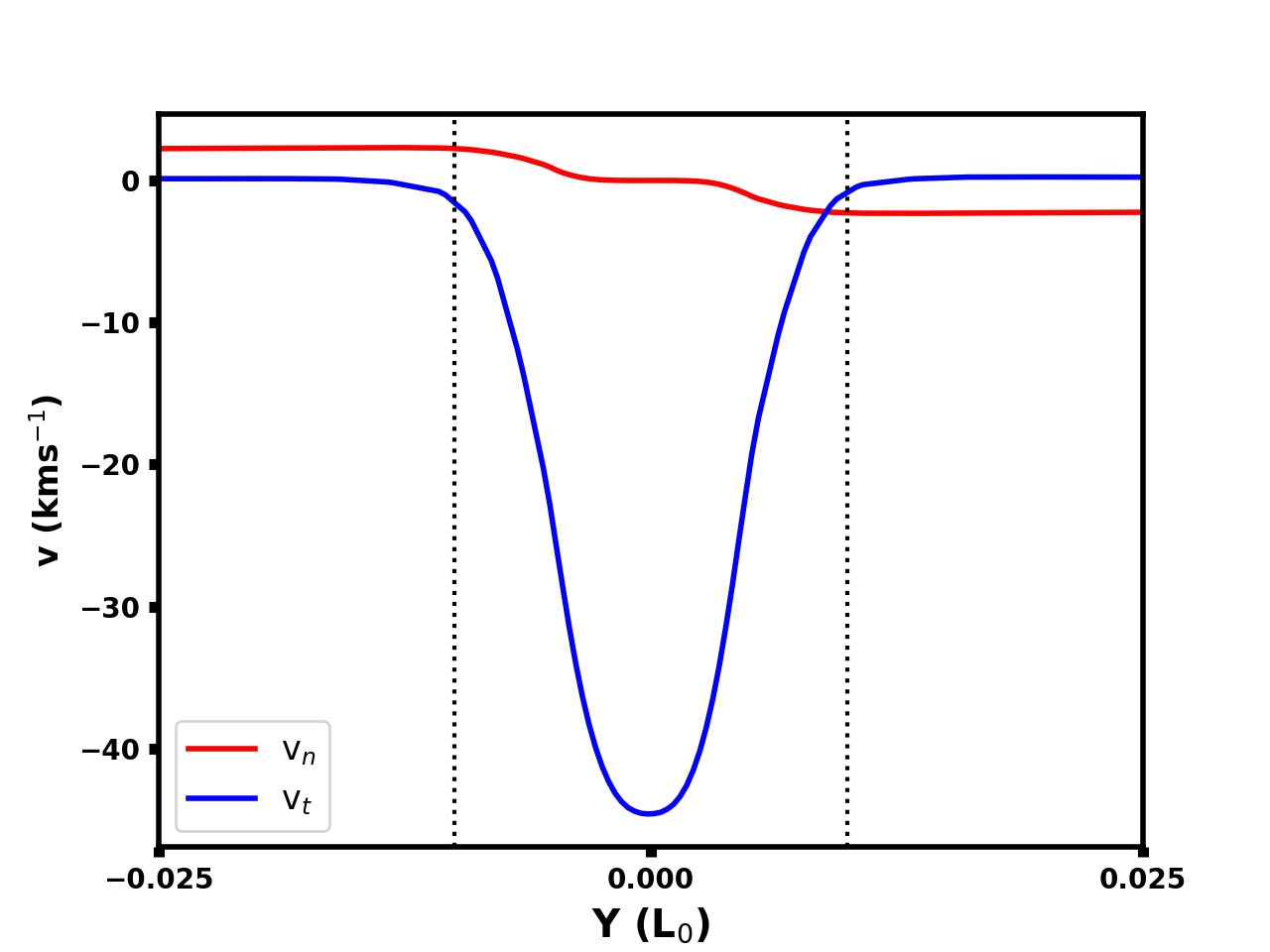}
\put(-33,23){\textbf{(g)}}
\end{minipage}

\caption{The distributions of magnetic field (solid lines) and out-of-plane electric current density j$_{z}$ at four different time points during the magnetic reconnection process in the case with $\beta_{0}$ = 0.05 and b$_{pert}$ = 0.005 at Z=1250 km above the solar surface (a), 
Distribution of the primary variables along the vertical line in the third panel of j$_{z}$ which crosses a pair of shocks (b)-(g). 
B$_{n}$ (and V$_{n}$) is the normal component of the magnetic field (and velocity), and B$_{t}$ (and V$_{t}$) is transverse component of the shock front, respectively.
The black vertical lines in (b)-(g) represent the positions of the shock front. The red boxes in (a) are the regions for calculating the reconnection rates.}
\label{fig_3} 
\end{figure*}


\section{Simulation results} \label{sec:results}

Figure~\ref{fig_1}(a) shows time evolution of the reconnection rate, $\gamma$ = V$_{y-aver}$/V$_{A-aver}$~\citep{zafar2023high}, for different initial perturbations (b$_{pert}$) at middle chromospheric altitudes. 
V$_{y-aver}$ and V$_{A-aver}$ are the average inflow and Alfven velocities within the red boxes in Figure~\ref{fig_3}(a).
For the reconnection rate calculation, we used boxes of different sizes near the principal reconnection X-point. We found that the box size affects the amplitudes of reconnection rates slightly but has no effect on the evolution trend. The main conclusion does not depend on the choices of the boxes used to calculate the reconnection rate.
The solid and dash-dotted lines correspond to Z = 1250 km and 1400 km, respectively.   
Like a typical reconnection process, an initial elongated Sweet-Parker phase with a reconnection rate of the order of $10^{-3}$ is followed by the faster plasmoid-mediated phase, where the reconnection rate reaches to around 0.02~\citep{ni2015fast}.  
Then another abrupt explosive reconnection process occurs, the reconnection rate is additionally increased to a higher value.
The maximum value reaches to around 0.06 or above in the explosive Petschek-like stage and then drop back to $\sim$0.02. 
The explosive Petschek-like phase is followed by a plateau phase, where the reconnection rate remains at its maximum value for a period of time. The plateau phase persists longer for small b$_{pert}$, but its duration decreases with increasing initial magnetic perturbation. The onset time of explosive reconnection is also affected by the strength of initial magnetic perturbation, which increases by decreasing b$_{pert}$.
The case with a high initial perturbation b$_{pert}$ = 0.03 (red line in Figure~\ref{fig_1}(a)) shows several sharp spikes in the reconnection rate. 
The initial spike in the reconnection rate right after the start of the simulation (red solid line in Figure 1(a)), where the reconnection rate reaches around 0.02, is due to the strong initial perturbation in this simulation case. 
However, the other tall short-lived spikes in the reconnection rate at t $\sim$ 14 sec and 16 sec correspond to explosive Petschek-like reconnection, indicating that this is an intermittent process.

Next, the effects of initial plasma $\beta$ ($\beta_{0}$) and plasma parameters at different altitudes (Table~\ref{tab:table1}) on the dynamics of this newly observed explosive fast intermittent reconnection events are explored. 
Figure~\ref{fig_1}(b) shows that an explosive Petschek-like reconnection phase occurs at both middle chromospheric altitudes, regardless of the initial plasma $\beta$ value.
Simulations with higher $\beta_{0}$ have a longer plateau phase, whereas those with lower $\beta_{0}$ have a shorter one.
The explosive reconnection phase is triggered earlier when strength of the initial magnetic field is stronger, and the initial plasma $\beta$ is lower.
The time evolution of the reconnection rate at various altitudes, from the photosphere to the top of the chromosphere with the same initial plasma $\beta$ ($\beta_{0}$ = 0.5) and perturbation (b$_{pert}$ = 0.009) shows that the reconnection rate peaks-up violently and attains the values of about 0.06 or above (Figure~\ref{fig_1}(c)).

The fast reconnection phase is observed in all simulation cases,
suggesting that the transition from plasmoid-dominated to the explosive faster Petschek-like reconnection phase are independent of magnetic perturbations (b$_{pert}$), initial plasma $\beta$ ($\beta_{0}$) and physical parameters ($T_{0}, \rho_{0}$) of various altitudes in the lower solar atmosphere. 
However, the difference in b$_{pert}$ and $\beta_{0}$ effect the duration of the plateau phase and the timing of the explosive reconnection regime onset. 
Here we want to mention that no explosive reconnection phase has been observed at Z = 2000 km in the presence of ambipolar diffusion with $\beta_{0} = 1.33$ and b$_{pert} = 0.005$ in our previous study~\citep{zafar2023high}. 
To clarify the dynamics of explosive faster reconnection phase at Z = 2000 km above the solar surface, we performed simulations with different initial plasma $\beta$ ($\beta_{0}$ =  0.3, 3.0 and 5.0) and initial magnetic perturbations. 
Additionally, we carried out a simulation with the same initial plasma $\beta$ ($\beta_{0} = 1.33$) as in Case-VI from the previous study, but with a perturbation of b$_{pert} = 0.001$. An explosive Petschek-like reconnection phase is observed in all these simulations with different $\beta_{0}$ and b$_{pert}$ at Z = 2000 km (not shown here).
We reviewed Case-VI (Z = 2000 km, $\beta_{0}$ = 1.33, b$_{pert} = 0.005$) from our previous study ~\citep{zafar2023high} and found that this simulation case terminated earlier, which could reveal why an explosive faster reconnection phase was not observed.
Therefore, we expect that the plasmoid-mediated phase in Case-VI of the previous study would have transformed into an explosive Petschek-like reconnection phase, if the simulations had been run for a longer duration.


To explore the underlying trigger mechanism of explosive fast reconnection, the temporal evolution of various plasma parameters are analyzed during the explosive fast reconnection stage 
in each case, Figure~\ref{fig_2}(a) shows the results in the case with $\beta_{0}=0.05$, $b_{pert}=0.005$ at Z=1250 km (green solid line in Figure~\ref{fig_1}(a)) above the solar surface. 
It is quite evident that the decrease in plasma temperature, density and pressure at the main X-point occurs just before the onset of explosive fast reconnection.   
Rapid drop in plasma temperature and density cause a reduction in gas pressure at the X-point, an increase in the plasma inflow velocity and leads to a spontaneous fast magnetic reconnection.
After careful analysis of the simulation results, we identified several reasons for the decrease in plasma density and temperature at the primary X-point. 
These include radiative cooling, motion of the primary X-point along the current sheet and ejection of hot plasma from the primary X-point.
Radiative cooling restricts the increase in plasma temperature during the reconnection process and has a significant impact on the distribution of plasma density and temperature.
In some cases, such as photospheric and lower chromospheric reconnection events, it has a substantial contribution to decrease plasma temperature.
In the non-linear stage, the distributions of plasma temperature and density along the current sheet is non-uniform and there is a possibility that the X-point moves from the hot and dense region to the cool and less dense one, leading to faster decrease in temperature and density. 
Another possibility is the ejection of hot plasma from the main X-point and its surroundings, resulting in a sudden decrease of plasma temperature, density and hence pressure.
Two drastic spikes in the reconnection rate at Z = 2000 km (see the cyan line in Figure~\ref{fig_1}(c)) at t = 85 sec and 105 sec corresponds to temperature drop caused by the motion of the X-point and the ejection of hot plasma from the surrounding of X-point, respectively.

The magnetic diffusion caused by electron-ion collision, $\eta_{ei}$  and electron-neutral collision, $\eta_{en}$ both increase significantly when the temperature at the X-point drops during t = 28.5 sec to t = 31.5 sec in the case with $\beta_{0} = 0.05$ and b$_{pert}$ = 0.005 at Z = 1250 km above the solar surface (Figure~\ref{fig_2}(b)).
The distributions of the logarithm of the total magnetic diffusion through the main X-point along the y-direction at different times are shown in Figure~\ref{fig_2}(c).
It is observed that the magnetic diffusion at the main X-point is actually much smaller than that of outside the current sheet. But, the local Petschek-like reconnection still happens.

Figure~\ref{fig_3}(a) shows the 2D distributions of the current density J$_{z}$ (background color) and field lines (black solid lines) at various stages of the reconnection process. 
After the Sweet-Parker reconnection stage, the current sheet become unstable and leads to multiple plasmoids. 
With the occurrence of the plasmoid instability the reconnection rate peaks up and reaches to a value of about 0.02.
One such stage having a chain of magnetic islands is shown in Figure~\ref{fig_3}a(i), at the same time the reconnection rate is about $\sim$ 0.02 (green line in Figure~\ref{fig_1}(a)).
Subsequently, all magnetic islands wiped out with time and the current sheet becomes bifurcated (Figures~\ref{fig_3}a(ii) and (iii)). 
In the proximity of X-point, the magnetic field lines have V-shaped structure while it possesses W-shaped structure when far away from the X-point. 
These are important characteristic of Petschek like scenario. 
The peak of magnetic reconnection (t = 28.5-31.5 s, green line in Figure~\ref{fig_1}(a)) corresponds to the phase where the current sheet dynamics is Petschek like.
At about t = 31.5 s, the current sheet again becomes unstable and breakup into plasmoids (Figure~\ref{fig_3}a(iv)). 
The reconnection rate drops again to around 0.02 (t$>$31.5 s, green line in Figure~\ref{fig_1}(a)) with the generation of these new plasmoids. 
The simulation with initial perturbation of 0.03 (red line in Figure~\ref{fig_1}(a)) clearly shows that this is a cyclic process, and the reconnection rate fluctuates between 0.07 and 0.02.


To examine Petschek shocks, the distributions of temperature $(T)$, mass density $(\rho)$, gas pressure $(P)$, out of plane current density $(J_{z})$, magnetic fields $(B)$ and velocity $(V)$ along the dotted black sampling line in Figure~\ref{fig_3}a(iii) are illustrated in Figures~\ref{fig_3}(b)-\ref{fig_3}(g).
It is evident that most variables changes drastically as it crosses the shock fronts marked by dotted lines in the Figures~\ref{fig_3}(b)-\ref{fig_3}(g). 
The absolute values of B$_{t}$ and V$_{n}$ decreases, B$_{n}$ does not vary much, and the other physical quantities increase from the upstream to the downstream regions.
Such discontinuous structure have been investigated in many previous studies~\citep{sato1979strong,ugai1979magnetic,mei2012numerical}, where the slow mode shocks are confirmed to be formed. 
This analysis reveals that the system changes from plasmoid-mediated phase to the Petschek-like reconnection (t $\sim$ 28.5s-31.5 green solid line in Figure~\ref{fig_1}(a)) and hence leading to explosive faster reconnection with the rate $\geq$ 0.06 becomes established. 


\section{Discussion and Conclusion} \label{sec:summary}
We have investigated the magnetic reconnection phenomena at various altitudes in the partially ionized low solar atmosphere with different initial conditions based on the single-fluid MHD model.
The time-dependent ionizations of the partially ionized helium-hydrogen plasmas and suitable radiative cooling models are considered in these numerical experiments.
The results presented in this work reported for the first time that the plasmoid-dominated fast reconnection phase make transition to an explosive faster Petschek-like reconnection phase in partially ionized plasmas.
 

The careful analysis has revealed that the sudden decrease of plasma temperature and density at the main X-point are responsible for sudden drop of plasma pressure, leading to an explosive fast reconnection rate.
The bifurcated shape of the local current sheet and existence of pair of slow mode shocks during the newly observed faster reconnection phase clearly proved the occurrence of the Petschek-like reconnection.
The reconnection rate reached to 0.06 or above during the explosive phase in all these simulation experiments, which is almost three times higher than that during the plasmoid-dominated reconnection phase.
The reconnection rate drops back to $\sim0.02$ when the local current sheet with the primary X-point gets elongated and new plasmoids are generated.
The occurrence of such a faster reconnection phase is independent of initial plasma $\beta$, magnetic perturbation and plasma parameters at different altitudes. However, the timing of the explosive reconnection regime onset, and the duration of the plateau phase is much smaller in the lower $\beta$ and higher perturbation cases.

The Petschek-like reconnection observed in our simulations differs from the classical Petschek reconnection model, that is derived based on the static state MHD equations. 
The classical one is a steady reconnection model, and the derived reconnection rate scales with Lundquist number S as $\gamma \sim$ 1/logS.
Therefore, the reconnection rate predicated by the classical Petschek model is 0.1 for a typical Lundquist number S $\sim10^{10}$ of a flare current sheet in the corona, which is sufficiently fast to explain the timescale of solar flares.
However, MHD simulations failed to obtain a Petschek solution for a uniform resistivity profile.
The Petschek solution can be physically realized only when there is a localized anomalous resistivity at the X-point.
It is worth nothing that in order to produce anomalous resistivity, the length scale of the current sheet is comparable to the ion Larmor radius or the ion inertial length, hence non-MHD (collisionless effects) are likely to become important before steady-state Petschek reconnection is realized~\citep{zweibel2009magnetic}.
In addition, such a steady reconnection process cannot well match the explosive reconnection process with bursty radiation enhancements. 
However, the Petschek-like reconnection in our simulations is non-steady process, highly dynamic, and intermittent in the whole reconnection process.


We should also mention that the unstable local Petschek-like reconnection has been identified during the plasmoid instability stage in the previous simulations with full ionized plasmas~\citep{baty2012onset,shibayama2015fast,shibayama2019mechanism}, 
but the reconnection rates in those works only reached about 0.01-0.02, the second faster reconnection stage was not presented. Our results revealed a fast reconnection mechanism with a reconnection rate of order 0.1 in the partially ionized low solar atmosphere in the MHD scale.
Apart from partially ionized solar plasma, our analysis can also be useful to better understand fast magnetic reconnection in other partially ionized environments such as interstellar medium, protoplanetary discs and magnetic reconnection studies in laboratory experiments. 

\section*{Acknowledgments}
This research was supported by the National Key R$\&$D Program of China Nos. 2022YFF0503804 (2022YFF0503800) and 2022YFF0503003 (2022YFF0503000); the NSFC Grants 12373060 and 11933009; the Strategic Priority Research Program of CAS with grants XDB0560000 and XDA17040507; the outstanding member of the Youth Innovation Promotion Association CAS (No. Y2021024); the Yunling Talent Project for the Youth; the Basic Research of Yunnan Province in China Grant 202401AS070044; the Yunling Scholar Project of the Yunnan Province and the Yunnan Province ScientistWorkshop of Solar Physics; Yunnan Key Laboratory of Solar Physics and Space Science under the number 202205AG070009; The numerical calculations and data analysis have been done on Hefei advanced computing center and on the Computational Solar Physics Laboratory of Yunnan Observatories. We benefit from the discussions of the ISSI-BJ Team "Solar eruptions: preparing for the next generation multi-waveband coronagraph."

\bibliography{sample631}{}
\bibliographystyle{aasjournal}



\end{document}